\documentclass[lettersize,journal,twocolumn]{IEEEtran}
\usepackage{graphicx}
\graphicspath{{Figures/}}
\usepackage{amsfonts}
\usepackage{amssymb}
\usepackage{amsmath}
\usepackage{color}
\usepackage{wasysym}
\usepackage{hyperref}
\usepackage{bbold}
\usepackage{algpseudocode}
\usepackage{placeins}
\usepackage{makecell}

\usepackage{mathtools}

\definecolor{lightblue}{RGB}{136,163,209}

\usepackage[numbers,sort&compress]{natbib}

\usepackage{algorithm}
\usepackage{algpseudocode}

\usepackage{authblk}

\begin{document}

\author[1]{Francesco Ferrari}
\affil[1]{Quantum Computing Solutions, Leonardo S.p.A., Via R. Pieragostini 80, Genova, 16151, Italy}
\author[1]{Matteo Vandelli}
\author[1,2]{Daniele Dragoni}
\affil[2]{Leonardo Hypercomputing Continuum, Leonardo S.p.A., Via R. Pieragostini 80, Genova, 16151, Italy}


\title{Feasibility-driven QAOA with penalty scheduling}

\maketitle

\begin{abstract}
Most available quantum algorithms address constrained optimization problems by treating constraints as soft penalty terms within a QUBO formulation. This approach requires careful adjustment of the penalty coefficients, which scales poorly with the number of constraints and lacks a proper strategy to balance feasibility and solution quality. In this work, we introduce two extensions of standard linear-ramp QAOA (lr-QAOA) tailored to problems with multiple heterogeneous constraints. We first construct $\Lambda$-lr-QAOA, in which each penalty term is assigned its own linear-ramp schedule, promoting penalty weights from external hyperparameters to internal variational parameters of QAOA, similarly to the objective and mixer parameters. By optimizing all schedules jointly in a single run, this approach eliminates nested penalty tuning and scales more efficiently to multiple constraints. The optimization is guided by a feasibility-driven loss function that pushes the quantum state towards high-quality \emph{feasible} solutions. As a further refinement, we introduce piecewise-ramp QAOA, in which the linear ramps are replaced by two-segment piecewise schedules, enhancing the expressiveness of the Ansatz at the cost of a small 
parameter overhead independent of the circuit depth. We benchmark both methods on Earth-observation satellite mission planning tasks formulated as budget-constrained Maximum Weight Independent Set problems. Numerical results show that piecewise-ramp QAOA consistently outperforms lr-QAOA and $\Lambda$-lr-QAOA across circuit depths and system sizes. Furthermore, both $\Lambda$-lr-QAOA and piecewise-ramp QAOA exhibit a high feasibility rate, which is crucial in industrial applications. Our analysis highlights an intrinsic feasibility-optimality trade-off, which we address by introducing a filtered variant of the loss providing a single hyperparameter to tune this balance.
\end{abstract}


\section{Introduction}

Combinatorial optimization problems arise in numerous industrial applications, such as scheduling, routing, facility placement and resource allocation~\cite{petropoulos2024or}. They typically involve the optimization of an objective function, either maximized or minimized, subject to a set of constraints~\cite{nemhauser1988integer,hillier2001introduction}. Many of these problems are NP-hard, meaning exact solution algorithms scale exponentially with the number of variables and become impractical for large instances~\cite{woeginger2003exact}. As an alternative, problem-specific heuristics~\cite{hochba1997approx} and approximate methods~\cite{kochenderfer2019algorithms} (e.g., simulated annealing, tabu search, genetic algorithms) are commonly used to find near-optimal solutions efficiently.

Alongside these methods, quantum computing approaches have emerged as a promising alternative, leveraging quantum effects such as superposition and tunneling to effectively explore complex solution landscapes~\cite{abbas2024review,egger2026quantum}. In this context, algorithms such as quantum annealing and variational methods are commonly employed. However, except for few specific cases, enforcing problem constraints exactly within these frameworks remains challenging~\cite{hadfield2019qaoa,baertschi2020grover,leipold2021constructing,xie2025grover,vandelli2025constraint,bucher2025qaoa,onah2026empirical,bucher2026constrained}. More often, constrained problems are reformulated as Quadratic Unconstrained Binary Optimization (QUBO) models, where constraints are incorporated into the objective function through soft penalty terms that disfavor infeasible configuration~\cite{kochenberger2014qubo,lucas2014ising,glover2019qubo}. In practice, however, this approach makes it difficult to guarantee the feasibility of solutions, particularly in realistic and large-scale applications, as selecting appropriate penalty coefficients is itself an NP-hard problem~\cite{alessandroni2025alleviating}.

\begin{figure*}[!t]
    \centering
    \includegraphics[width=\textwidth]{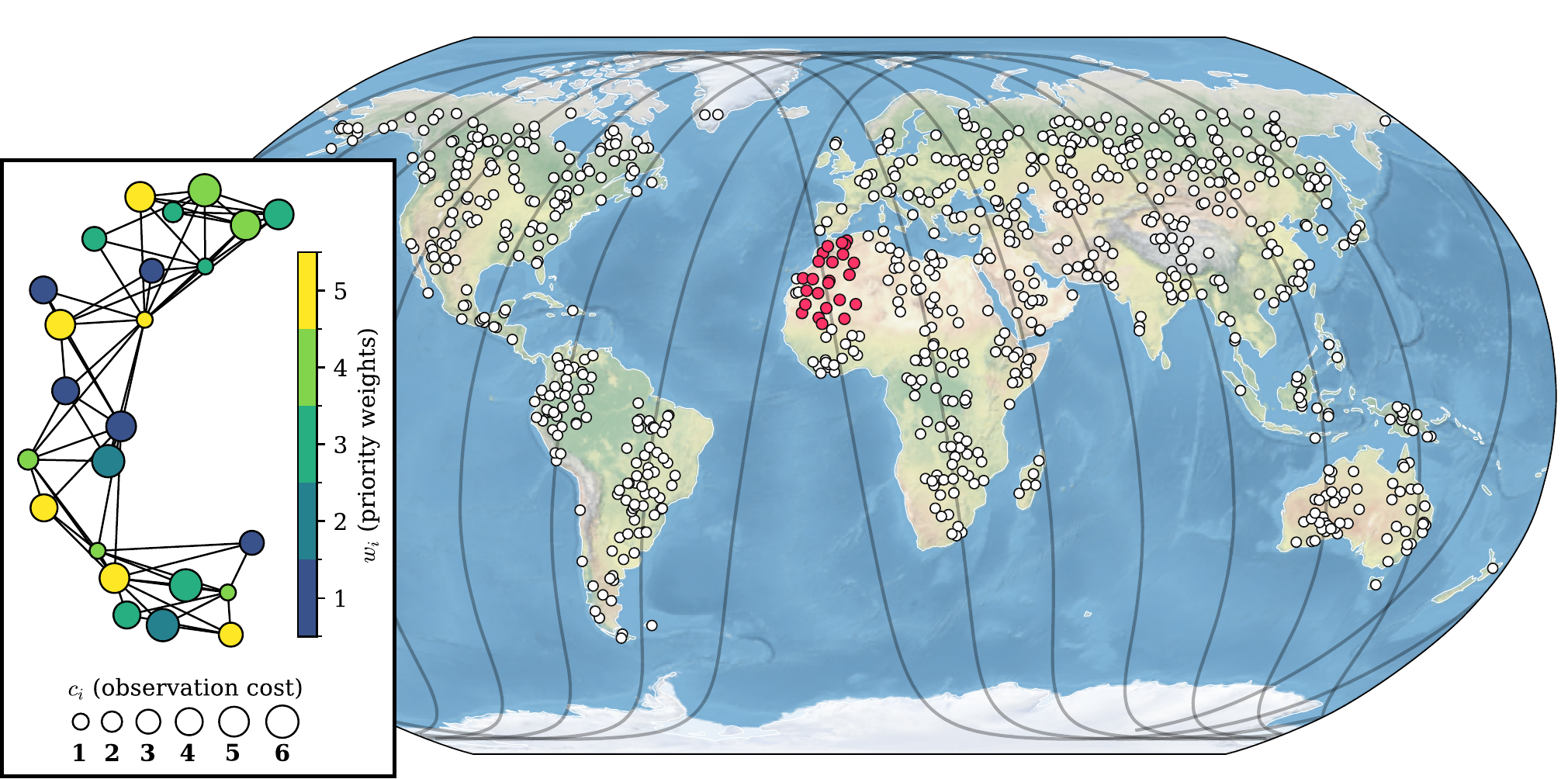}
    \caption{Satellite mission planning problem. Dots on the map denote observation targets and gray lines represent the satellite ground track. The red points mark a disconnected subset of $24$ targets that define the budget-constrained MWIS~\eqref{eq:mwis_obj} instance analyzed in the numerical experiments. The inset shows the corresponding conflict graph, in which an edge between two targets indicates that they cannot be scheduled jointly. Node colors represent the priority weights $w_i$ assigned to each target, while node sizes represent the corresponding observation costs $c_i$. For this instance the total available budget is $B=20$.}
    \label{fig:sat_instance}
\end{figure*}

Nevertheless, several quantum computing approaches have demonstrated promising results even within the QUBO framework. Among variational quantum methods, the Quantum Approximate Optimization Algorithm (QAOA)~\cite{farhi2014qaoa} has become a primary focus for near-term applications, offering a flexible framework for tackling combinatorial optimization~\cite{blekos2024qaoa}. Yet, scaling QAOA to higher circuit depths introduces a rapid growth in the dimensionality of the variational parameter space, making optimization increasingly challenging~\cite{bittel2021training,shaydulin2019multistart,zhou2020qaoa,larocca2022barren,vandelli2024evaluating,kossmann2025deepqaoa}. To address this issue, various methods for reducing the number of variational parameters (or transfer them across instances) have been explored~\cite{brandao2018fixed,willsch2022gpu,barraza2022piecewise,sureshbabu2024parameter,montanezbarrera2025transfer,saavedrapino2025qaoa,sakai2025transferring,he2026regularized,mcdowall2026spectral,guo2026transfer}, among which the linear-ramp QAOA (lr-QAOA) stands out as a simple yet empirically successful strategy~\cite{mbeng2019quantumannealing,shaydulin2021classical,kremenetski2021quantum,sack2021quantum,kremenetski2023quantum,he2025parameter,dehn2025extrapolation}, as demonstrated by recent extensive benchmarks~\cite{montanezbarrera2025lrqaoa}. By constraining the parameters to follow a linear schedule, this approach simplifies the optimization process and improves practicality while still yielding high-quality solutions.

In this work, we propose an extension of lr-QAOA motivated by the practical requirements of real-world constrained optimization problems, where obtaining high-quality \emph{feasible} solutions is often as valuable as finding the actual global optimum. The method, which we term \emph{piecewise-ramp QAOA}, relies on two key ideas. The first is that penalty terms encoding problem constraints are not treated as fixed coefficients in a static QUBO formulation, as is standard, but are instead assigned their own independent variational schedules within the QAOA ansatz. This allows the algorithm to jointly optimize solution quality and constraint satisfaction in a single run, eliminating the need for penalty tuning and naturally extending to problems with multiple constraints by assigning variational parameters to each one of them. The loss function is defined as the expectation value of the objective Hamiltonian evaluated over \emph{feasible bitstrings only}, which drives the algorithm toward high-quality feasible solutions. The second key idea is to replace the single linear ramp of lr-QAOA with a two-segment piecewise linear schedule for the objective and penalty terms. This generalization adds one breakpoint per schedule, modestly increasing the number of variational parameters while allowing the ansatz to explore a broader class of adiabatic trajectories.

The proposed method is numerically tested through emulation. As a case study, we consider a minimal toy model of satellite mission planning~\cite{mitrovic2019planning,chen2019milp,wang2021aeos,li2025review}, in which observation tasks need to be optimally scheduled based on their priority and cost; this leads to a budget-constrained Maximum Weight Independent Set (MWIS) problem~\cite{gabrel2003mathematical,duncan2021mwis,stollenwerk2021aeos,quetschlich2023hybrid,makarov2024quantum,delilbasic2024reverse,delilbasic2026quantum}

\section{Problem}

The optimal scheduling of satellite observation tasks can be formulated as a combinatorial optimization problem, and has been the subject of several recent studies exploring the applicability of quantum optimization methods. Various formulations of the problem, mostly recast in QUBO form, have been tackled using quantum~\cite{stollenwerk2021aeos,makarov2024quantum,casati2025exploiting,marchioli2025quantum,baioletti2026applying} and hybrid~\cite{quetschlich2023hybrid,rainjonneau2023quantum,delilbasic2024reverse,delilbasic2026quantum} algorithms. Here we consider a minimal model for a satellite mission planning problem in which the task is to schedule a set of ground target acquisitions to be performed by an agile satellite within a given time horizon. Each candidate observation opportunity corresponds to a ground target that becomes visible during a satellite pass and is represented as a node $i\in V$ of a conflict graph $G=(V,E)$. An edge $(i,j) \in E$ is introduced whenever two observation opportunities $i$ and $j$ are incompatible, i.e. when they cannot be jointly scheduled within their respective visibility windows for any feasible choice of timing, accounting for the satellite slew time between targets and a post-reorientation settling period. Within this setting, an independent set of the graph $G$ corresponds to a set of observations that admits a feasible joint schedule. Each observation is assigned a certain level of priority, measured by the positive weight $w_i$, and a cost $c_i>0$, which can reflect different aspects of the observation burden, such as the data storage capacity required by the acquisition or the onboard energy required by the imaging payload. The aim of the mission planning task is to select a feasible subset of mutually compatible observations that maximizes the total priority, while keeping the total cost within a resource budget $B$. The problem is then modeled as a budget-constrained Maximum Weight Independent Set (MWIS) problem 
\begin{align}
    & \max_{x} \sum_{i \in V} w_i x_i \label{eq:mwis_obj} \\
\text{s.t.} 
& \quad x_i + x_j \leq 1 \quad \forall (i,j) \in E, \label{eq:is_const} \\
& \ \  \sum_{i\in V} c_i x_i \leq B,  \label{eq:budget_const} \\
& \quad x_i \in \{0,1\} \quad \forall i \in V. \nonumber
\end{align}
An illustrative instance of this problem is shown in Fig.~\ref{fig:sat_instance}. The binary variables $x_i$ take the value $1$ when the corresponding target $i$ is selected for acquisition. The \emph{local} independent set (IS) constraint of Eq.~\eqref{eq:is_const} ensures that no pairs of incompatible observations are scheduled simultaneously, while the \emph{global} budget constraint of Eq.~\eqref{eq:budget_const} enforces that the total cost is kept within the assigned budget. This formulation can be interpreted as a preliminary selection layer within a more complex scheduling pipeline, identifying a resource-feasible subset of mutually compatible observations to be passed to a detailed scheduler for precise timing within their visibility windows. As a deliberately simplified model of a full satellite mission planning task, it ignores several operational aspects; the primary purpose is to assess the performance of the proposed quantum algorithm on a combinatorial instance of industrial relevance, derived from a realistic operational scenario. 

We note that the budget-constrained MWIS problem outlined above, also known as the Knapsack Problem with Conflict Graph, is NP-hard for general instances~\cite{garey1979computers,bettinelli2017knapsack}, motivating the development of alternative computational approaches such as quantum optimization algorithms.

\section{Methods}

\subsection{QUBO formulation}

Tackling binary combinatorial optimization problems with quantum algorithms usually involves the reformulation of the problem in terms of an Ising spin model, mapping the binary variables into Pauli $Z$-operators, i.e. $x_i \mapsto (1-Z_i)/2$. The objective function is typically transformed into an objective Hamiltonian, $H_{\text{obj}}$, whose ground state corresponds to the optimal solution of the problem, provided that constraints are satisfied. For instance, the MWIS objective function~\eqref{eq:mwis_obj} yields $H_{\text{obj}}=\sum_i w_i Z_i$. However, as previously noted, the exact enforcement of constraints within quantum optimization algorithms remains a significant challenge.

The most common approach is to handle them approximately via a QUBO formulation~\cite{kochenberger2014qubo,lucas2014ising,glover2019qubo}, in which each constraint is incorporated as a penalty term added to the objective Hamiltonian, designed to raise the energy of the bitstrings that violate the constraint, while leaving feasible states unaffected. The total QUBO Hamiltonian is then expressed as 
\begin{equation}\label{eq:H_QUBO}
{H_{\rm QUBO} = H_{\text{obj}} + \sum_k \lambda_k H^{(k)}_{\text{pen}}},    
\end{equation} 
where the label $k$ indicates the various Hamiltonian penalty terms $H^{(k)}_{\text{pen}}$. The coefficients $\lambda_k >0$ are chosen to ensure that infeasible states are energetically penalized. For example, for the IS constraint of Eq.\eqref{eq:is_const}, one can take a term $x_ix_j$ for each edge, which translates to the penalty Hamiltonian ${H^{(\rm IS)}_{\text{pen}} = \sum_{(i,j)\in E} (1-Z_i)(1-Z_j)}$; for the budget constraint of Eq.\eqref{eq:budget_const}, the standard approach is to introduce auxiliary slack variables to convert the inequality into an equality, which is then penalized through a quadratic Hamiltonian. Here, instead, we adopt the unbalanced penalization scheme of Ref.~\cite{montanezbarrera2024unbalanced}, which avoids the use of slack variables. Within this scheme, the penalty consists of the sum of a linear term, ${H^{(\rm B1)}_{\text{pen}} = \left[\sum_{i \in V} c_i (\frac{1-Z_i}{2}) -B\right]}$, and a quadratic term, ${H^{(\rm B2)}_{\text{pen}} = \left[\sum_{i \in V} c_i (\frac{1-Z_i}{2}) -B\right]^2}$, each with its own penalty coefficient.

We note that, in general, the choice of suitable penalty coefficients $\lambda_k$ is not straightforward as it involves a delicate trade-off between feasibility and optimality. If $\lambda_k$ are too small, the quantum solver may favor an infeasible state that yields a low value of the objective function. Conversely, if $\lambda_k$ are too large, the energy landscape becomes dominated by the penalty terms, which mask the original objective function and can create a rugged optimization surface. Indeed, finding optimal values of $\lambda_k$ is generally NP-hard~\cite{alessandroni2025alleviating}.

\subsection{Standard linear-ramp QAOA for QUBO}

The lr-QAOA method combines aspects of variational quantum algorithms and quantum adiabatic approaches~\cite{montanezbarrera2025lrqaoa}. The system evolves from an initial Hamiltonian $H_{\rm mix}$, whose ground state $|\psi_{\rm mix}\rangle$ is easy to prepare, towards the problem Hamiltonian $H_{\rm QUBO}$ encoding the QUBO instance. The time evolution is approximated by $p$ Trotter steps of alternating unitaries, in which the contribution of each Hamiltonian follows a discretized linear ramp
\begin{equation}\label{eq:U_lr-qaoa}
    U_{\rm lr}(\beta, \gamma) = \prod_{l=1}^{p} e^{-i \beta (1-\frac{l}{p}) H_{\rm mix}}  e^{-i \gamma\frac{l}{p} H_{\rm QUBO}},
\end{equation}
Here, the $\beta$ and $\gamma$ parameters are optimized variationally in order to minimize the QUBO energy expectation value over the final state $|\psi_{\rm lr}\rangle= U_{\rm lr}(\beta, \gamma)|\psi_{\rm mix}\rangle $, i.e.
\begin{equation}
    \mathcal{L}_{\rm QUBO}=\langle\psi_{\rm lr} | H_{\rm QUBO} | \psi_{\rm lr}\rangle = \sum_{\sigma}  E_{\rm QUBO}(\sigma)  \, |\langle \sigma | \psi_{\rm lr} \rangle|^2,
\end{equation}
where $\sigma$ runs over all bitstrings and $E_{\rm QUBO}(\sigma)=\langle \sigma | H_{\rm obj} | \sigma \rangle$. This scheme provides a physically motivated ansatz that approximates the continuous adiabatic schedule in a Trotterized, gate-based setting.

However, in practical applications, alongside the optimization of the schedule parameters $\beta$ and $\gamma$, a careful tuning of the $\lambda_k$ penalty coefficients is often required to drive the algorithm towards feasible solutions. In this sense, the $\lambda_k$ penalties play the role of hyperparameters of the QUBO-based formulation of lr-QAOA. While problem-specific bounds allow sometimes to choose $\lambda_k$ analytically~\cite{lucas2014ising}, no principled criterion exists for arbitrary constraints. Pre-computation strategies have been proposed to determine suitable values of 
$\lambda_k$ in advance, although their practicality can be strongly problem-dependent~\cite{alessandroni2026scalable}. In the absence of such strategies, the most common approach is to perform a grid search over candidate values of $\lambda_k$, running the full circuit optimization for each choice and selecting the value that best balances solution quality and feasibility. Overall, this results in a nested optimization loop in which the outer level searches over penalty coefficients and the inner level optimizes the circuit parameters, significantly increasing the total computational overhead compared to unconstrained problems. This motivates the search for alternative, more direct penalty optimization strategies.

\begin{figure}[!t]
    \centering    \includegraphics[width=0.85\columnwidth]{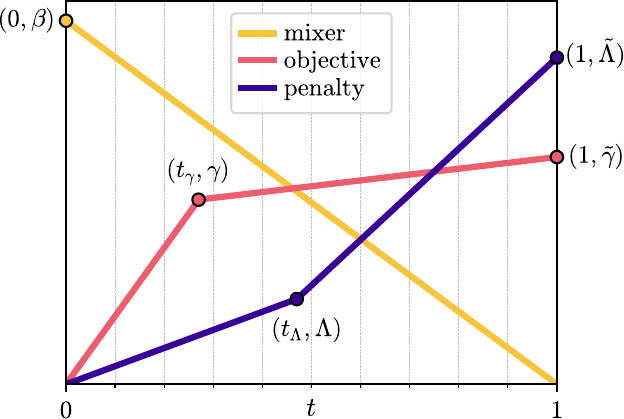}
    \caption{Illustrative sketch of the piecewise-ramp QAOA algorithm, showing the time schedules of mixer, objective and penalty Hamiltonians (assuming a single penalty term is present). Vertical lines indicate the discretization of the ramps over $p=10$ Trotter steps.}
    \label{fig:piecewiseQAOA}
\end{figure}

\subsection{Variational penalty optimization}

We introduce a modified version of lr-QAOA in which the objective Hamiltonian $H_{\rm obj}$ and each penalty term $H^{(k)}_{\rm pen}$ are assigned \emph{independent} linear ramps, rather than being combined into a single QUBO Hamiltonian as in Eq.~\eqref{eq:H_QUBO}. As a result, the penalty coefficients $\lambda_k$ are promoted from fixed hyperparameters to genuine variational parameters, denoted as $\Lambda_k$ and optimized along with $\beta$ and $\gamma$. We refer to this approach as $\Lambda$-lr-QAOA. The corresponding time-evolution operator reads
\begin{align}\label{eq:U_feas_lr-qaoa}
    U_{\Lambda\text{-lr}}(&\beta, \gamma, \{\Lambda_k\})  \nonumber \\
    & = \prod_{l=1}^{p} \left[ e^{-i \beta (1-\frac{l}{p}) H_{\rm mix}}  e^{-i \gamma\frac{l}{p} H_{\rm obj} } \prod_k e^{-i \Lambda_k \frac{l}{p} H^{(k)}_{\rm pen}} \right].
\end{align}
Once penalties are treated variationally, the QUBO energy ceases to be a meaningful loss function and a different one must be introduced. Motivated by the demands of real-world applications, where high-quality feasible solutions retain practical value even when the global optimum is not reached, we define the \emph{feasibility-driven} loss function
\begin{equation}\label{eq:feas-cost}
    \mathcal{L}_{F}=\sum_{\sigma\in F}  E_{\rm obj}(\sigma)  \, |\langle \sigma | \psi \rangle|^2.
\end{equation}
Here, $|\psi\rangle$ denotes the output state of the variational circuit. For the $\Lambda$-lr-QAOA circuit, this is explicitly given by ${|\psi_{\Lambda\text{-lr}}\rangle=U_{\Lambda\text{-lr}}(\beta, \gamma, \{\Lambda_k\})|\psi_{\rm mix}\rangle}$. The sum in Eq.~\eqref{eq:feas-cost} runs only over the set of feasible bitstrings $F$, effectively assigning zero energy to infeasible ones. For this to be meaningful, i.e. for $\mathcal{L}_{F}$ to favor feasible over infeasible states, the objective energy must satisfy $E_{\rm obj}(\sigma)=\langle \sigma | H_{\rm obj} | \sigma \rangle < 0$ for all $\sigma\in F$. This holds, for instance, in the MWIS problem of Eq.~\eqref{eq:mwis_obj}. More generally, for a broad class of problems one can enforce a slightly stronger condition, namely that the objective Hamiltonian is negative on \emph{all} bitstrings: it is enough to subtract a constant offset larger than $\max_\sigma E_{\rm obj}(\sigma)$. For binary problems with a linear objective function, this maximum is efficiently computable. For quadratic objectives, a simple upper bound on the maximum energy is given by the sum of the absolute values of all coefficients.

Under this condition, $\mathcal{L}_{F}$ is minimized by concentrating probability on low-energy feasible bitstrings. In the limit of an exact solution, where the final state of $\Lambda$-lr-QAOA is sharply peaked on the optimal bitstring(s), $\mathcal{L}_{F}$ reduces to the actual ground state energy of the objective Hamiltonian $H_{\rm obj}$. We note also that $\mathcal{L}_{F}$ is suited to hardware implementations, as it can be estimated from a finite set of measurement outcomes by accumulating the objective energy over sampled bitstrings that satisfy the constraints, discarding the rest.

Overall, the $\Lambda$-lr-QAOA simultaneously tunes the weight of the objective function ($\gamma$) and the penalties ($\Lambda_k$) within a single optimization run, circumventing the need for a nested optimization. The advantage becomes particularly significant when multiple heterogeneous constraints are present: each introduces an independent $\Lambda_k$, and jointly optimizing all of them avoids the prohibitive multi-dimensional hyperparameter grid search required in the standard QUBO-based lr-QAOA approach.

\begin{figure*}[!t]
    \centering
    \includegraphics[width=\textwidth]{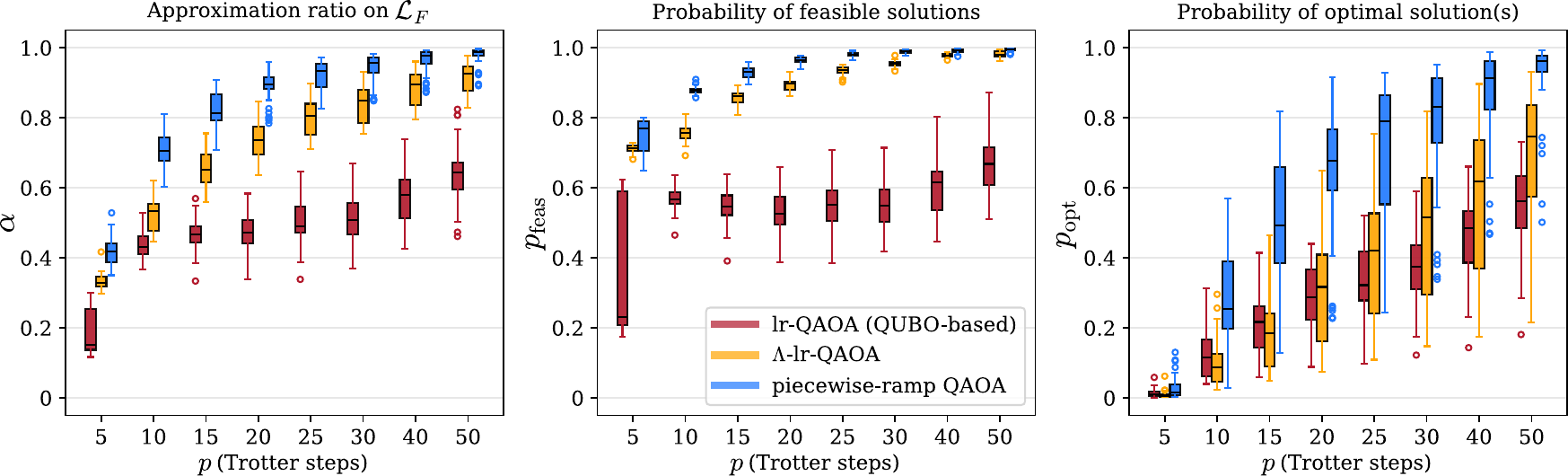}
    \caption{Benchmark results on the MIS problem. The three panels contain box plots of $\alpha$, $p_{\rm feas}$ and $p_{\rm opt}$ as a function of the number of Trotter steps $p$. For each value of $p$, we collect statistics over $30$ MIS instances defined on Erd\H{o}s--R\'enyi graphs with $|V|=18$ nodes. 
  The results of piecewise-ramp QAOA (blue) are compared to the standard QUBO-based lr-QAOA (red) and to $\Lambda$-lr-QAOA (orange).}
    \label{fig:mis_vs_p}
\end{figure*}

\subsection{Piecewise-ramp QAOA}

The adoption of the loss function of Eq.~\eqref{eq:feas-cost} allows for further extensions of $\Lambda$-lr-QAOA, with arbitrary time schedules. In principle, one could setup a full QAOA with distinct ($\beta, \gamma, \{\Lambda_k\}$) parameters for each layer. However, this quickly becomes challenging for large circuit depths, as the number of variational parameters grows linearly with $p$, leading to a high-dimensional and difficult optimization landscape~\cite{bittel2021training,kossmann2025deepqaoa}. We choose instead to introduce a \textit{piecewise-ramp QAOA} approach to increase the expressiveness of the Ansatz while keeping the parameter count low. The corresponding time-evolution operator is
\begin{align}\label{eq:U_piecewiseQAOA}
    U_{\rm pw}(\beta; t_\gamma, \gamma,  \tilde{\gamma}; & \{t_{\Lambda_k}, \Lambda_k, \tilde{\Lambda}_k\}) \nonumber \\ 
    = \prod_{l=1}^{p} & \left[ 
    \vphantom{\prod_k e^{-i  f_{\rm pw}(\frac{l}{p}, t_{\Lambda_k}, \Lambda_k, \tilde{\Lambda}_k)  H^{(k)}_{\rm pen}}}
    e^{-i\beta(1-\frac{l}{p}) H_{\rm mix}}  e^{-i f_{\rm pw}(\frac{l}{p}; t_\gamma, \gamma, \tilde{\gamma}) H_{\rm obj} } \right. \nonumber \\ 
    & \qquad \qquad \left.
    \prod_k e^{-i  f_{\rm pw}(\frac{l}{p}; t_{\Lambda_k}, \Lambda_k, \tilde{\Lambda}_k)  H^{(k)}_{\rm pen}} \right],
\end{align}
where
\begin{equation}
f_{\rm pw}(t; t_a, a, \tilde{a}) = \begin{cases}
a\dfrac{t}{t_a} & \text{if } 0 \le t \le t_a \\[2mm]
a + \dfrac{\tilde{a} - a}{1 - t_a} \, (t - t_a) & \text{if } t_a < t \le 1
\end{cases}
\end{equation}
Here, the mixer Hamiltonian is progressively turned off following a linear ramp, as in ($\Lambda$-)lr-QAOA. Instead, for the objective and penalty Hamiltonians, two-step piecewise ramps $f_{\rm pw}$ are used, in which the breakpoint coordinates (time and value) and the final value are optimized. For instance, for $H_{\rm obj}$, the ramp goes from $(0,0)$ to the breakpoint $(t_\gamma, \gamma)$, and then from there to the final value $(1,\tilde{\gamma})$. The various ramps are Trotterized over $p$ steps, i.e. taking discrete times $t=l/p$, as indicated in Eq.~\eqref{eq:U_piecewiseQAOA}. The piecewise-ramp QAOA thus relies on $4+3N_{\rm pen}$ variational parameters, with $N_{\rm pen}$ the number of distinct penalty terms. We note that the parameter count is independent of the number of Trotter steps $p$, in the same spirit as lr-QAOA. Furthermore, for a given value of $p$, the circuit depth is the same as lr-QAOA, since the piecewise ramp only modifies the time schedules of the Hamiltonians, without introducing additional layers or gates. A representative illustration of the schedules of piecewise-ramp QAOA is given in Fig.~\ref{fig:piecewiseQAOA} for the case with a single penalty term. The parameters of piecewise-ramp QAOA are optimized minimizing the loss $\mathcal{L}_{F}$ in Eq.\eqref{eq:feas-cost} over the output state $|\psi_{\rm pw}\rangle=U_{\rm pw}(\beta; t_\gamma, \gamma,  \tilde{\gamma}; \{t_{\Lambda_k}, \Lambda_k, \tilde{\Lambda}_k\})|\psi_{\rm mix}\rangle$.

\section{Benchmarks on simplified problem instances}

\subsection{Numerical details}

To conduct extensive benchmarks across varying problem sizes, we first consider a simplified setting in which node weights are assumed to be uniform and the available budget is large enough not to constrain the solution (i.e., $B \geq \sum_{i\in V} c_i$). This reduces the problem to the unweighted Maximum Independent Set (MIS) problem, which is still NP-hard~\cite{garey1979computers} and serves as a well-established benchmark for combinatorial optimization. Multiple instances are constructed by generating random Erd\H{o}s--R\'enyi graphs with edge probability $0.5$, a challenging regime for MIS due to the absence of exploitable structure. If a sampled graph is not connected, we enforce connectivity by adding randomly selected edges between disconnected components until the graph becomes connected. 

We compare the results of piecewise-ramp QAOA (with $7$ variational parameters) with those obtained using the standard lr-QAOA with $H_{\rm QUBO}$ as objective function and $\lambda=1.1$, which, for the MIS problem, is sufficiently large to make violations of the IS constraint energetically unfavorable~\cite{lucas2014ising}. Futhermore, we also consider the $\Lambda$-lr-QAOA of Eq.~\eqref{eq:U_feas_lr-qaoa} in which the penalty weight for the IS constraint $\Lambda_{\rm IS}$ is treated as a variational parameter.

For the optimization of the variational parameters of the different methods we use the (gradient-free) differential evolution algorithm~\cite{storn1997de}, followed by a L-BFGS-B run~\cite{byrd1995BFGS} to refine the best solution in the final population. We note that, for numerical stability, it is convenient to renormalize the Hamiltonians in the circuits of Eqs.~\eqref{eq:U_lr-qaoa}, \eqref{eq:U_feas_lr-qaoa} and~\eqref{eq:U_piecewiseQAOA} dividing each of them by the largest absolute coefficient in their respective expansions over Pauli strings. Furthermore, the time parameters of the breakpoints $t_a$ are obtained by applying a sigmoid function to the corresponding unconstrained variational parameter to ensure that they remain within the interval $[0,1]$. 

The metrics used for the comparison of the different methods are the probability of sampling feasible solutions, $p_{\rm feas}$, and the probability of sampling optimal solutions, $p_{\rm opt}$. Furthermore, we compute the approximation ratio $\alpha$ between the loss function of Eq.~\eqref{eq:feas-cost} and its value at the optimal solution, which corresponds to the energy of the optimal bitstring(s). The upper-bound $\alpha=1$ indicates that the final wave function is fully peaked on the optimal solution(s) of the constrained problem, while smaller values denote deviations from optimality. 

The different quantum algorithms have been implemented using the \emph{Qiskit} library from IBM~\cite{Qiskit}. Numerical emulations of the quantum circuits have been run on our proprietary \emph{davinci} supercomputer, equipped with Intel 8568 Emerald Rapids $48$-Core CPUs and NVIDIA H200 GPUs. 

\begin{figure*}[!t]
    \centering
    \includegraphics[width=\textwidth]{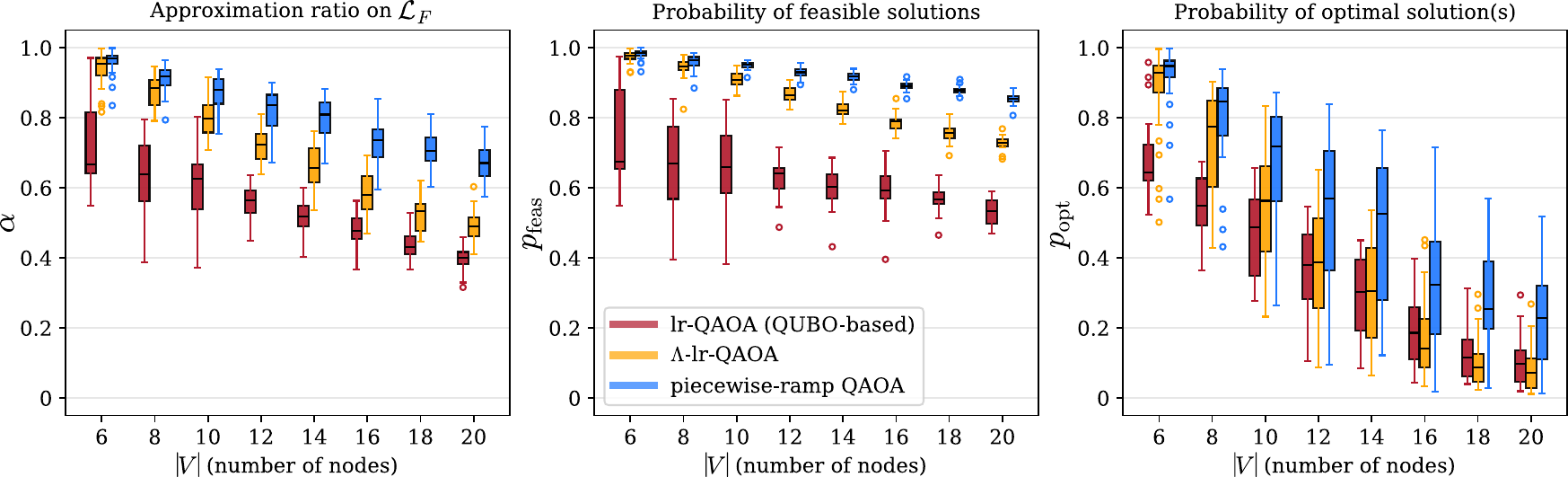}
    \caption{Benchmark results on the MIS problem. The three panels contain box plots of $\alpha$, $p_{\rm feas}$ and $p_{\rm opt}$ as a function of the number of graph nodes $|V|$. For each value of $|V|$, we collect statistics over $30$ MIS instances on Erd\H{o}s--R\'enyi graphs. Calculations are performed with $p=20$ Trotter steps. The results of piecewise-ramp QAOA (blue) are compared to the standard QUBO-based lr-QAOA (red) and to $\Lambda$-lr-QAOA (orange).}
    \label{fig:mis_vs_N}
\end{figure*}

\subsection{Results}

We begin by considering a pool of $30$ MIS instances on graphs with $|V|=18$ nodes to study the dependence of $\alpha$, $p_{\rm feas}$ and $p_{\rm opt}$ as a function of the number of Trotter steps $p$. The results are summarized in the boxplots of Fig.~\ref{fig:mis_vs_p}, where the same pool of instances is used at the various values of $p$. Overall, the quality of the results improves with $p$, as expected. The $\Lambda$-lr-QAOA approach yields final states with significantly higher probability of sampling feasible bitstrings compared to the standard QUBO-based lr-QAOA, reaching values close to $1$ at large circuit depths. The probability of finding the optimal solution is comparable between the two methods, with the standard lr-QAOA performing slightly better at shallow depths ($p\leq 15$) and $\Lambda$-lr-QAOA taking over beyond that. Most importantly, the piecewise-ramp QAOA consistently outperforms the other two methods across all values of $p$, demonstrating that the additional expressiveness of the piecewise schedule leads to a better simultaneous optimization of solution quality and constraint satisfaction. In particular, $p_{\rm opt}$ displays a much faster growth as a function of the number of Trotter steps, getting very close to $1$ at $p=50$.

A similar analysis applies to the boxplots of Fig.~\ref{fig:mis_vs_N}, which reports results across graphs of varying size $|V|$ (between $6$ and $20$ nodes) at fixed number of Trotter steps $p=20$. Here too, piecewise-ramp QAOA consistently yields feasible bitstrings with high probability across the whole range of sizes considered here, closely followed by $\Lambda$-lr-QAOA, whose performance deteriorates more rapidly as $|V|$ increases. Moreover, piecewise-ramp QAOA maintains a median $\alpha$ consistently above $0.6$ across all system sizes, displaying only a slow decrease with $|V|$. For what concerns $p_{\rm opt}$, while the median of piecewise-ramp QAOA always stays above $0.2$, both lr-QAOA and $\Lambda$-lr-QAOA show a faster decay, with the latter outperforming the former at small $|V|$, but both approaching similar performance at large system sizes.

\begin{figure}[!t]
    \centering
    \includegraphics[width=0.8\columnwidth]{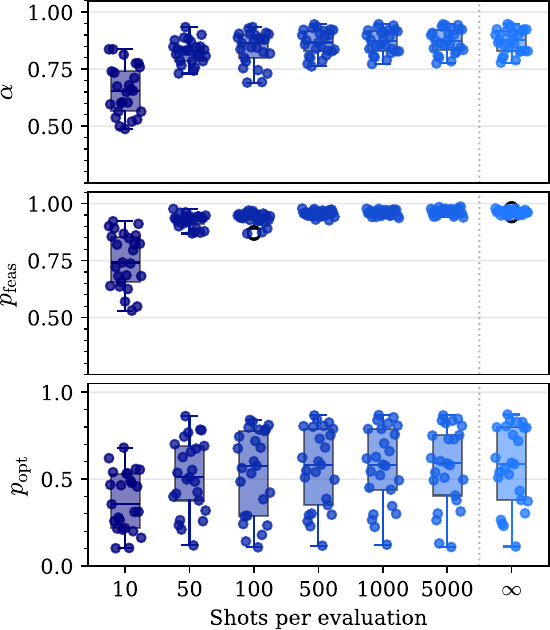}
    \caption{Benchmark results of the piecewise-ramp QAOA on the MIS problem. Results are obtained on a set of $25$ instances of $|V|=18$ nodes, using $p=20$ Trotter steps.    The three panels contain box plots of $\alpha$, $p_{\rm feas}$ and $p_{\rm opt}$ as a function of the number of shots used during optimization to estimate the loss function, simulating the finite-sampling regime of actual quantum devices.}
    \label{fig:mis_vs_shots}
\end{figure}

To conclude our benchmark analysis on MIS, we simulate the finite-sampling regime of real quantum hardware by running the piecewise-ramp QAOA optimization with a finite number of shots per circuit execution to estimate the loss function. The final quality metrics are instead evaluated exactly via statevector simulation. The results, displayed in Fig.~\ref{fig:mis_vs_shots}, show that the pipeline is rather stable and the results converge quite fast as a function of the number of shots. This robustness is likely due to the noise tolerance of differential evolution and to the unbiasedness of the shot-based estimator of $\mathcal{L}_{F}$. We expect, however, that larger system sizes would require more shots, as the output distribution becomes more spread and individual bitstrings carry less weight.

\section{Satellite mission planning}

\subsection{Problem instances}

We now move beyond synthetic benchmark cases and consider more realistic instances of the satellite mission planning problem defined in Eq.~\eqref{eq:mwis_obj}. To generate concrete instances we consider the IRIDE-MS2-HEO 1 satellite, a Low Earth Orbit (LEO) satellite, part of the IRIDE Earth observation constellation designed for environmental monitoring and disaster management~\cite{iride}. We propagate the satellite orbit over a time window of $12$ hours (starting from a random date in 2026), using publicly available Two-Line Element data and the \emph{skyfield} Python library~\cite{skyfield}. A set of $800$ candidate observation targets is randomly placed on land near the satellite ground track by applying random spatial displacements, as shown in Fig.~\eqref{fig:sat_instance}. In order to avoid polar orbital artifacts, targets are constrained to lie within latitudes $[-65^\circ,65^\circ]$. A target $i$ is considered observable during a pass if the satellite elevation angle from the target is larger than $20^\circ$. For each target, we keep only the best observation opportunity, defined as the one achieving the highest peak elevation angle. 

To determine incompatibilities between observation tasks, we make the following simplifying assumptions. We assume that each observation requires a dwell time of $\tau=150$ seconds and that the satellite has a slew velocity of $v=0.5^\circ/s$ to reorient between consecutive observations. Furthermore, we assume that an additional settling time of $T=30$ seconds is needed after each satellite reorientation. Therefore, given two observation opportunities $i$ and $j$, the minimum transition time required between them is $\Delta_{ij} = d(i,j)/v + T$, where $d(i,j)$ is the slew angle between the targets.

Two observations $i$ and $j$, with visibility time windows $[a_i,b_i]$ and $[a_j, b_j]$, are compatible if they can both be executed within their respective visibility windows in at least one of the two possible orderings. Let us assume that $i$ is performed before $j$: the best case is obtained by scheduling $i$ as early as possible, finishing at time $a_i+\tau$, and scheduling $j$ as late as possible, starting at time $b_j-\tau$. The ordering is feasible if the available time gap is large enough, i.e. if $(b_j-\tau)-(a_i+\tau) \geq \Delta_{i,j}$. The same check is repeated for the reverse ordering and the pair is declared incompatible, i.e. the edge $(i,j)$ is added to the conflict graph, if and only if both orderings are infeasible. 

Within this scheme, the conflict graph $G$ resulting from the full set of $800$ candidate observations always consists of several disconnected components of varying size, reflecting the inherent sparsity of the conflicts induced by the orbital geometry and the finite visibility windows. For our numerical calculations, we select isolated subgraphs of moderate size that remain tractable for classical emulation of the quantum algorithms.

\subsection{Feasibility vs Optimality}

\begin{figure}[!t]
    \centering
    \includegraphics[width=0.9\columnwidth]{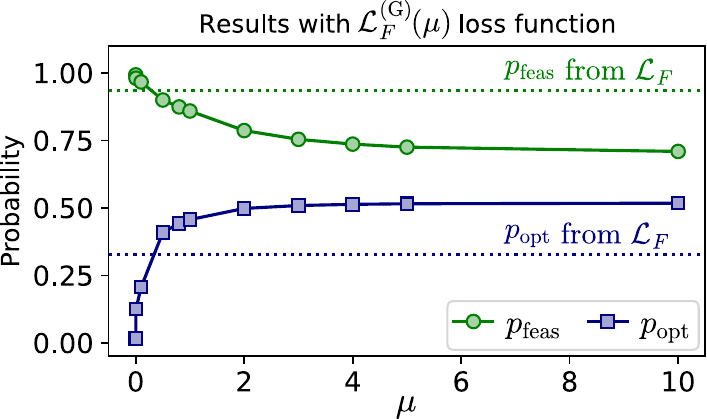}
    \caption{Probabilities to sample feasible and optimal bitstrings from the final state of piecewise-ramp QAOA (satellite mission planning instance of Fig.~\ref{fig:sat_instance}; $p=20$ Trotter steps). Results are obtained using the filtered loss function of Eq.~\eqref{eq:feas-cost-filter} and plotted as a function of $\mu$, with green circles and blue squares representing $p_{\rm feas}$ and $p_{\rm opt}$, respectively. The horizontal lines represent the corresponding values achieved by the piecewise-ramp QAOA based on the (unfiltered) loss $\mathcal{L}_{F}$.}
    \label{fig:prob_vs_mu}
\end{figure}

Let us consider the satellite mission planning scenario depicted in Fig.~\eqref{fig:sat_instance} and focus on the isolated subset of $24$ observation tasks highlighted by the red points. Each observation $i$ is assigned a random integer priority weight $w_i$, drawn uniformly from $\{1, 2, \dots, 5\}$. Observation costs $c_i$ are drawn uniformly  from $\{1, 2, \dots, 6\}$, and the total budget is set to $B=20$. The resulting conflict graph representing the budget-constrained MWIS problem is shown in the inset of Fig.~\eqref{fig:sat_instance}. This instance has $4987$ feasible bitstrings out of $16.7$ million, with three degenerate optimal solutions at $E_{\rm obj}=-25$. Tackling it with the QUBO-based lr-QAOA requires careful tuning of three penalty coefficients, for which we performed an expensive search over an iteratively refined three-dimensional grid. The best results in terms of optimality over the parameter grid, using $p=20$ Trotter steps, are $p_{\rm opt}\approx 0.20$, $\alpha \approx0.44$ and $p_{\rm feas} \approx0.49$. In contrast, a single optimization run of piecewise-ramp QAOA with the same circuit depth yields a noticeably improved solution, with $p_{\rm opt}\approx 0.33$, $\alpha \approx 0.85$ and a total feasibility of $p_{\rm feas}\approx 0.94$.

Focusing on the $24$-node instance under consideration, we examine the inherent tension between feasibility and optimality that arises in constrained optimization problems. Within standard QUBO-based frameworks, calibrating penalty coefficients to simultaneously maximize the probability of sampling optimal and feasible solutions is notoriously difficult. The choice of the loss function $\mathcal{L}_{F}$ is designed to strike a balance between these two tendencies, steering the wave function weight toward the low-energy feasible subspace, rather than concentrating it on the optimal bitstring(s). Alternative formulations of the loss function are possible. In particular, we introduce a variant of $\mathcal{L}_{F}$ that allows one to control the feasibility-optimality trade-off through a single hyperparameter $\mu\in \mathbb{R}_+$, by applying an exponential filter to the bitstring energies~\cite{amaro2022filtering}. The resulting filtered loss function takes the form of a Gibbs free energy~\cite{li2020gibbs}
\begin{equation}\label{eq:feas-cost-filter}
    \mathcal{L}^{\rm (G)}_F(\mu)=-\frac{1}{\mu} \log\left(\sum_{\sigma \in F} e^{-\mu E_{\rm obj}(\sigma)}  \, |\langle \sigma | \psi \rangle|^2 \right).
\end{equation}
We emphasize that also here the sum runs only over feasible bitstrings $F$. For large values of $\mu$, the weight assigned to the lowest-energy (feasible) bitstrings by $\mathcal{L}^{\rm (G)}_F(\mu)$ increases, and thus the tendency towards optimality. Conversely, as $\mu$ decreases, the influence of $E_{\rm obj}$ is progressively reduced; in this limit, the distribution favors the feasible subspace in a more unbiased fashion, prioritizing constraint satisfaction over cost minimization~\cite{footnote1}.

\begin{figure}[!t]
    \centering
    \includegraphics[width=\columnwidth]{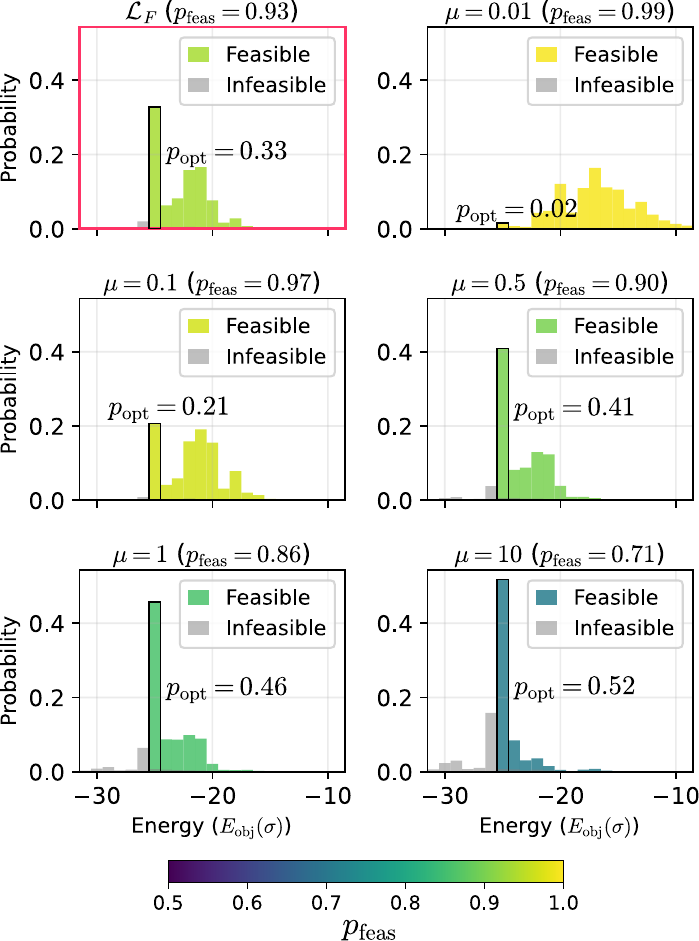}
    \caption{Distribution of bitstring probabilities ($|\langle \sigma | \psi_{\rm pw} \rangle|^2$) as a function of their objective energy $E_{obj}(\sigma)$. Results obtained by piecewise-ramp QAOA (with $p=20$) on the $24$-node satellite mission planning instance of Fig.~\ref{fig:sat_instance}. The upper-left panel, highlighted in red, is obtained by minimizing the unfiltered loss $\mathcal{L}_{F}$; the remaining panels are obtained with the filtered loss $\mathcal{L}^{\rm (G)}_F(\mu)$ of Eq.~\eqref{eq:feas-cost-filter}, at increasing values of $\mu$. Grey bars represent bitstrings violating the constraints; colored bars represent feasible bitstrings, with the color indicating total $p_{\rm feas}$. The bar at the optimal energy is highlighted by a black contour.}
    \label{fig:histo}
\end{figure}

We tackle the $24$-nodes instance of Fig.~\eqref{fig:sat_instance}, optimizing the piecewise-ramp QAOA to minimize the filtered loss function $\mathcal{L}^{\rm (G)}_F(\mu)$ at different values of $\mu$. The resulting feasibility and optimality probabilities are shown in Fig.~\ref{fig:prob_vs_mu}. Increasing the value of $\mu$ it is possible to push $p_{\rm opt}$ above $50\%$, at the cost of a reduced feasibility probability, which decreases to roughly $70\%$. On the contrary, when $\mu$ takes small values, feasibility is enhanced and gets very close to $1$, while optimality drops to small values. This illustrates how the single hyperparameter $\mu$ provides a practical knob to navigate the feasibility-optimality trade-off.

It is instructive to take a look at the distribution of the bitstring probabilities $|\langle \sigma | \psi_{\rm pw} \rangle|^2$. This is illustrated in Fig.~\ref{fig:histo} for the optimal piecewise-ramp QAOA states obtained by minimizing $\mathcal{L}_{F}$, and $\mathcal{L}^{\rm (G)}_F(\mu)$ at different $\mu$ values. At very small $\mu$, the distribution is broadly spread over many feasible bitstrings with negligible weight on infeasible ones, but the probability assigned to the optimal solutions remains small. As $\mu$ increases, the distribution visibly concentrates towards lower energies and the bar at the optimal energy grows. Concurrently, the probabilities of sampling infeasible bitstrings, shown in grey, increase. These distributions provide a visual illustration of the feasibility-optimality trade-off controlled by $\mu$.

\begin{figure}[!t]
    \centering
    \includegraphics[width=0.85\columnwidth]{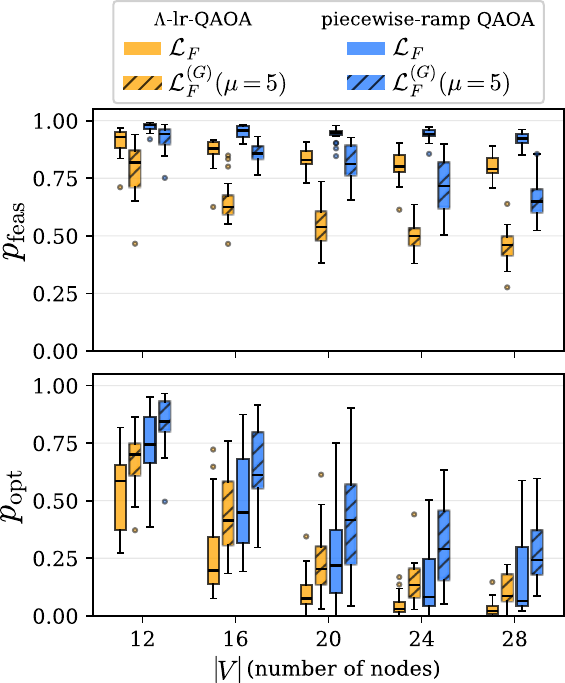}
    \caption{Results for the satellite mission planning problem~\eqref{eq:mwis_obj}. Boxplots of $p_{\rm feas}$ (top panel) and $p_{\rm opt}$ (bottom panel) are plotted as a function of the number of targets $|V|$, collecting statistics over $20$ instances per size. The results of piecewise-ramp QAOA (blue) are compared to those of $\Lambda$-lr-QAOA (orange), both employing $p=20$ Trotter steps. Plain and hatched boxplots indicate the different loss functions used for optimization, $\mathcal{L}_F$ and ${\mathcal{L}^{(G)}_F(\mu=5)}$, respectively.} 
    \label{fig:boxplots_sat}
\end{figure}

Finally, we perform calculations on a set of instances from the satellite mission planning problem with different number of targets $|V|$. We repeat the orbital propagation described above starting from random dates and take the resulting disconnected subgraphs, until we obtain $20$ instances per size $|V|\in \{12,16,20,24,28\}$. Running lr-QAOA for all these instances is prohibitive, as calibrating three penalty weights for each instance implies a substantial computational cost. $\Lambda$-lr-QAOA and piecewise-ramp QAOA, instead, require only a single optimization. A comparison of the results of two methods is shown in Fig.\ref{fig:boxplots_sat}. We used both the plain loss function $\mathcal{L}_F$ and its filtered version $\mathcal{L}^{\rm (G)}_F(\mu)$ at $\mu=5$. Using the former, piecewise-ramp QAOA achieves large percentages of feasibility, displaying a slower decay with $|V|$ than $\Lambda$-lr-QAOA. The optimality results show a wider spread, reflecting a substantial variability across instances. We observe how the filtered loss function is capable of consistently pushing $p_{\rm opt}$ to higher values, both for $\Lambda$-lr-QAOA and piecewise-ramp QAOA.

\section{Conclusion}

In this work, motivated by practical challenges of real-world constrained optimization, we introduced two variations of the regular linear-ramp QAOA approach~\cite{montanezbarrera2025lrqaoa} to target high-quality feasible solutions of constrained optimization problems, without increasing the circuit complexity of the original method. While finding the global optimum remains the ideal goal, obtaining near-optimal \emph{feasible} solutions is often of great practical value in industrial settings where constraint satisfaction is a strict requirement. Standard QUBO-based approaches often struggle to reliably meet this need, as the calibration of penalty coefficients involves a delicate trade-off between constraint satisfaction and solution quality.

The methods proposed in this work are two extensions of linear-ramp QAOA along two complementary directions. First, we promote the penalty coefficients in standard QUBO-based QAOA to additional variational parameters to be optimized. In this approach, dubbed $\Lambda$-lr-QAOA, each penalty Hamiltonian is assigned its own variational linear schedule, enabling finer control over the enforcement of constraints. To optimize these schedules, we use the feasibility-driven loss function of Eq.~\eqref{eq:feas-cost} that evaluates the objective function exclusively over feasible bitstrings, steering the algorithm toward near-optimal feasible solutions, without requiring explicit penalty calibration. Several advantages emerge from this combination. From a methodological standpoint, penalty coefficients are no longer treated as external hyperparameters that require a nested optimization loop or an expensive multi-dimensional grid search, but become instead internal variational parameters trained within a \emph{single} QAOA optimization run (analogously to the mixer and objective Hamiltonian parameters). This makes the approach particularly advantageous for problems with multiple heterogeneous constraints, where the standard QUBO recipe becomes impractical due to large dimensionality of the penalty space.

In addition to the promotion of penalty coefficients to variational parameters, we also introduced the \textit{piecewise-ramp QAOA}, in which the linear schedules of $\Lambda$-lr-QAOA are replaced by two-segment piecewise schedules for both the objective and penalty terms. This approach enhances the expressiveness of the $\Lambda$-lr-QAOA Ansatz at the cost of a small increase of the number of variational parameters to be optimized. From the standpoint of resource cost, the increase in parameter count is independent of the Trotter depth $p$ and the gate depth remains identical to that of linear-ramp QAOA at fixed $p$.

Both $\Lambda$-lr-QAOA and piecewise-ramp QAOA have been tested on Earth-observation satellite mission planning tasks, modeled as budget-constrained Maximum Weight Independent Set problems. We first benchmarked both approaches on simplified random instances of the problem, which effectively reduces to the MIS problem on random graphs. In these instances, piecewise-ramp QAOA consistently outperformed both the standard QUBO-based lr-QAOA and $\Lambda$-lr-QAOA across a wide range of circuit depths and problem sizes, achieving larger feasibility probabilities and a substantially faster growth of $p_{\rm opt}$ with the number of Trotter steps. Furthermore, our numerical tests show that the optimization remains rather stable when a shot-based estimator of the loss function is used, mimicking the finite-sampling regime of real quantum hardware.

We then turned to analyzing the performance of the algorithm on more realistic instances of the satellite mission planning task. For a sample instance of $24$ targets, a single optimization run of piecewise-ramp QAOA produced higher-quality solutions than those obtained from a computationally expensive three-dimensional grid search over the penalty coefficients of lr-QAOA. This demonstrates the tangible advantage of treating penalties variationally, particularly in the presence of multiple constraints. Finally, we introduced a filtered variant of the loss function, taking the form of a Gibbs free energy controlled by a single hyperparameter, which provides a simple mechanism to navigate the feasibility-optimality trade-off inherent to constrained problems. Increasing this hyperparameter concentrates the wavefunction weight on lower-energy solutions, at the cost of a reduced probability of sampling feasible states.

Several directions remain open for future work. The piecewise schedule could be extended to a richer class of parametrized ramps, balancing additional expressiveness against optimization complexity. Furthermore, execution on actual quantum hardware would be a possible next step toward assessing the scalability of the approach on problems of industrial interest.

\section*{Disclaimer}
The authors declare no competing interest. 

\section*{Acknowledgments}
The authors would like to thank the whole Quantum Computing Solutions group at Leonardo S.p.A. for useful discussions.

\bibliographystyle{unsrturl}
\bibliography{biblio}

\end{document}